\documentclass[aps,prb,twocolumn,amsmath,amssymb,superscriptaddress,floatfix]{revtex4}
\usepackage{graphicx}
\usepackage{bm}
\bibstyle{apsrev.bib}

\newcommand{\be}{\begin{equation}}
\newcommand{\ee}{\end{equation}}
\newcommand{\beqn}{\begin{eqnarray}}
\newcommand{\eeqn}{\end{eqnarray}}

\begin{document}

\title{Scaling properties at the interface between different critical subsystems: The Ashkin-Teller model}

\author{P\'eter Lajk\'o}
 \affiliation{Department of Physics, Kuwait University, P.O. Box 5969, Safat 13060, Kuwait}
\author{Lo\"{\i}c Turban}
 \email{turban@lpm.uhp-nancy.fr}
 \affiliation{Laboratoire de Physique des Mat\'eriaux, Universit\'e Henri 
Poincar\'e, BP~239,\\
F-54506 Vand\oe uvre l\`es Nancy Cedex, France}
\author{Ferenc Igl\'oi}%
 \email{igloi@szfki.hu}
 \affiliation{Research Institute for Solid State Physics and Optics,
H-1525 Budapest, P.O.Box 49, Hungary}
 \affiliation{Institute of Theoretical Physics,
Szeged University, H-6720 Szeged, Hungary}

\date{\today}

\begin{abstract}
We consider two critical semi-infinite subsystems with different critical exponents
and couple them through their surfaces. The critical behavior at the interface, influenced by
the critical fluctuations of the two subsystems, can be quite rich.
In order to examine the various possibilities, we study a system composed of two coupled Ashkin-Teller models with different four-spin couplings $\epsilon$,  on the two sides of the junction. By varying
$\epsilon$, some  bulk and surface critical exponents of the two subsystems are continuously
modified, which in turn changes  the interface critical behavior. In particular
we study the marginal situation, for which magnetic critical exponents  at  the interface
vary continuously with the strength of the interaction parameter. The behavior expected from scaling arguments is checked by density matrix renormalization group calculations.
\end{abstract}

\pacs{05.50.+q, 75.40.Cx, 75.70.Cn}

\maketitle

\section{Introduction}
\label{sec:intro}
Realistic systems have a finite extent and, when they display a second-order phase trans\-ition, the
critical behavior in the boundary region is generally different from that in the bulk.
The characteristic size of this region is given by the correlation length, which becomes divergent
as the critical temperature $T_c$  is approached. In the vicinity of the critical point, the
singularities of   local quantities, such as the surface magnetization, are characterized
by  critical exponents,  which are generally different from their bulk counterparts.
This type of  local critical phenomena has been thoroughly studied
in the case of a free surface through exact, field-theoretical and numerical
methods.\cite{binder83,diehl86,pleimling04}

When a system is in contact through its boundary with another system, the environment
can influence the local critical behavior at the interface. If, however, the critical temperature of the
environment, $T_c'$, is  different from $T_c$,
the nature of the transitions at the interface is expected to be the same as for a surface.\cite{berche91}  If the environment has the higher critical temperature $T_c' > T_c$,
it stays ordered at $T_c$ and the interface transition has the same properties as the extraordinary surface transition.\cite{binder83,diehl86,pleimling04}  In the opposite case, for $T_c' < T_c$, the
environment is disordered at $T_c$ and the interface transition is actually an ordinary surface transition.\cite{binder83,diehl86,pleimling04}

Here we consider the more complex problem when the two subsystems in contact have the same critical temperature but not the same set of critical exponents. Thus, the competition between the two different bulk and surface critical behaviors may result in
a completely new type of interface critical phenomena. This  problem has already been addressed
in Ref.~\onlinecite{bti06} in which the analytical mean-field solution, in terms of $\varphi^k$
field theories,  has been obtained and generalized by using phenomenological scaling considerations.
Monte Carlo simulations have also been performed in two dimensions  for interfaces between subsystems belonging to the universality classes of the Ising model, the three-state and four-state Potts models.

In all these examples the stable fixed points are related to surface critical behavior
and the expected renormalization group (RG) phase diagram is the one given in the upper part of Fig.~\ref{fig:phase2}.
For weak interface couplings, the junction renormalizes to a cut, and
we have the same local critical behavior as for a free surface, whereas for strong  
couplings,  the interface becomes ordered at the bulk transition temperature. For some 
intermediate value of the  couplings, there is a special interface transition fixed point,  involving new critical exponents,  which, however,  can be expressed in terms of the bulk and surface exponents of the
two subsystems.\cite{bti06}

In the present work  our purpose is  to examine the different  types of  possible interface
critical behavior which can be realized. Thus, we consider situations  where a weak interface
coupling can be  irrelevant, relevant or even truly marginal. We are particularly interested in the latter case.
A convenient system, for which all these different situations can be realized, is the two-dimensional (2D) Ashkin-Teller (AT)
model,\cite{AT} or its one-dimensional (1D) quantum version.\cite{KdNK81,IS84}

By introducing two Ising variables per site, the  AT Hamiltonian  can be rewritten as two Ising Hamiltonians coupled through a  four-spin interaction,\cite{fan72} which is a truly marginal operator.
As a consequence, some  bulk and  surface critical exponents  are continuously
varying functions of the strength of the four-spin coupling $\epsilon$.
These critical exponents are known exactly through conformal
invariance\cite{cardy87} and Coulomb-gas mapping.\cite{nienhuis87}

The composite system which we consider consists of two AT models with  the same critical
temperature but  different four-spin couplings, and thus different sets of critical exponents.
We couple these subsystems  through their surface spins and study the critical properties at  the interface  while varying the strength of the interface coupling. We first classify the possible interface critical
behaviors through scaling considerations,  which are then confronted with the results of extensive
numerical calculations using the density matrix renormalization group (DMRG).

The structure of the paper is as follows. The AT model and its basic
critical properties are described in Sec.~\ref{sec:AT}. We define the composite system and
discuss its possible interface RG phase diagrams in Sec.~\ref{sec:RG}. Results of numerical calculations are presented in Sec.~\ref{sec:num} and discussed in Sec.~\ref{sec:disc}.

\section{Ashkin-Teller model and its critical properties}
\label{sec:AT}
The AT model is defined in terms of two sets of Ising spin variables 
$\sigma_i=\pm 1$ and $\tau_i=\pm 1$, attached to each lattice site $i$.
The usual Ising interaction  $K(\sigma_i \sigma_j + \tau_i \tau_j)$ between nearest-neighbor sites $i$ and $j$  is supplemented by a four-spin interaction  $ K_4\sigma_i \sigma_j \tau_i \tau_j$, which is parametrized as $K_4=K\epsilon$. This latter term represents the product of the energy densities in
the two Ising systems. We consider the system on a square lattice
and work with the row-to-row  transfer matrix  $\cal T$.
In the Hamiltonian limit, the transfer matrix  can be written as ${\cal T} \sim \exp(-\kappa{\cal H}_{AT})$,
where ${\kappa}$ is the lattice spacing in the ``time"  direction and ${\cal H}_{AT}$ is the 1D quantum Hamiltonian given by
\beqn
{\cal H}_{AT}=&-&\sum_{i=1}^{L-1}(\sigma_i^z \sigma_{i+1}^z + \tau_i^z \tau_{i+1}^z)
-h \sum_{i=1}^{L}(\sigma_i^x+\tau_i^x)\cr
&-&\epsilon\left[\sum_{i=1}^{L-1}\sigma_i^z \sigma_{i+1}^z\tau_i^z \tau_{i+1}^z
+h \sum_{i=1}^{L} \sigma_i^x \tau_i^x \right]\,.
\label{H_AT}
\eeqn
Here, $\sigma_i^{x,z}$ and $\tau_i^{x,z}$ are two sets of Pauli matrices and $h$
is the strength of the transverse field, which plays the role of the temperature
in the classical system. One can introduce a set of dual Pauli operators 
$\widetilde{\sigma}_{i+1/2}^{x,z}$ and $\widetilde{\tau}_{i+1/2}^{x,z}$ such that 
\beqn
\widetilde{\sigma}_{i+1/2}^{x}=\sigma_{i}^{z}\sigma_{i+1}^{z}\,,
\quad \sigma_{i}^{x}=\widetilde{\sigma}_{i-1/2}^{z}\widetilde{\sigma}_{i+1/2}^{z}\cr
\widetilde{\tau}_{i+1/2}^{x}=\tau_{i}^{z}\tau_{i+1}^{z}\,,
\quad \tau_{i}^{x}=\widetilde{\tau}_{i-1/2}^{z}\widetilde{\tau}_{i+1/2}^{z}\,.
\label{dual}
\eeqn
When the Hamiltonian in Eq.~(\ref{H_AT}) is rewritten in terms of the dual variables, the couplings 
$J_i=1$ and the transverse fields $h_i=h$ exchange their roles.  Consequently, the homogeneous system 
is self-dual and the self-duality line is located at $h_c=1$. For $-1/\sqrt{2} \le \epsilon \le 1$ this is just
the critical line separating the ferromagnetic and the paramagnetic phases of the system. In the region
$-1<\epsilon \le -1/\sqrt{2}$,  for $h_c^{-}(\epsilon)<h<h_c^{+}(\epsilon)$, there is a so-called critical fan  in which the system stays critical.\cite{KdNK81}
At the critical point, the  excitation energy $\Delta E$ and the wave vector $k$ are linearly
related, $\Delta E=v_s k$, and the sound velocity is given by\cite{BvGR87_1}
\beqn
v_s=\frac{\pi \sin(\arccos\epsilon)}{\arccos\epsilon}\,.
\label{vs}
\eeqn

In the critical system, the basic operators are the magnetization ${\cal O}_m(i)=\sigma_i^z$
($\tau_i^z$), the energy density  ${\cal O}_e(i)=\sigma_i^z \sigma_{i+1}^z$ ($\tau_i^z \tau_{i+1}^z$)
or, through duality, $\sigma_i^x$ ($\tau_i^x$), and the polarization ${\cal O}_p(i)=\sigma_i^z \tau_{i}^z$.
The connected critical correlation functions display a power-law decay, so that
$\langle {\cal O}_{\alpha}(i){\cal O}_{\alpha}(i+r) \rangle- \langle {\cal O}_{\alpha}(i)\rangle\langle{\cal O}_{\alpha}(i+r) \rangle\sim r^{-2x_{\alpha}}$,
where $x_{\alpha}$ is the anomalous dimension of  ${\cal O}_{\alpha}$.
Similarly, surface-to-surface  correlations
involve the corresponding surface dimensions  $x_{\alpha}^s$.

The critical properties of the AT model are exactly known
through conformal invariance and Coulomb-gas mapping. The anomalous dimensions of bulk
operators are given by\cite{KdNK81}
\be
x_m=\frac{1}{8}\,,\quad x_e=\frac{\pi}{2 \arccos(-\epsilon)}\,,\quad x_p=\frac{1}{4} x_e\,.
\label{bulk-exp}
\ee
The correlation length critical exponent is related to the dimension of the energy density by $\nu=1/(2-x_e)$ when $-1/\sqrt{2} \le \epsilon \le 1$, whereas it is formally infinite  in the critical fan. True marginal behavior implies that the scaling dimension of the operator ${\cal O}_4(i)$ associated with the four-spin interaction $\sigma_i^z \sigma_{i+1}^z\tau_i^z \tau_{i+1}^z$  keeps the constant value  $x_4=2$, the same as for the two decoupled Ising chains.

The corresponding anomalous dimensions for surface operators are\cite{vGR87}
\be
x_m^s=\frac{\arccos(-\epsilon)}{\pi}\,,\quad x_e^s=2\,,\quad x_p^s=1\,.
\label{surf-exp}
\ee
One may notice that the anomalous dimensions, which are $\epsilon$ dependent in the bulk due to the presence of the marginal four-spin interactions, remain constant at the surface and  vice versa.

\section{Composite system and  renormalization group phase diagrams}
\label{sec:RG}
\subsection{Ladder and chain junctions}
A composite AT system is obtained by coupling two different semi-infinite  subsystems through their surface spins. These subsystems have  the same nearest-neighbor coupling, thus the same critical temperature. They have different values of the four-spin couplings $\epsilon^{(-)}$ ($\epsilon^{(+)}$) for $z<0$ ($z>0$) with $\epsilon^{(-)} \le \epsilon^{(+)}$. 

The junction can be of two different kinds, ladder or chain junction\cite{bariev79}
(see Fig.~6.1 of Ref.~\onlinecite{ipt93}). In the ladder junction, there are nearest-neighbor
as well as four-spin couplings between sites at $i=-1$ (boundary of the $z<0$ subsystem)
and $i=1$ (boundary of the $z>0$ subsystem). In the
Hamiltonian limit,  this corresponds to a term 
\be
{\cal V}_{-1,1}=-J(\sigma_{-1}^z\sigma_{1}^z + \tau_{-1}^z\tau_{1}^z +
\epsilon_{\rm int} \sigma_{-1}^z\sigma_{1}^z \tau_{-1}^z\tau_{1}^z)\,,
\label{V}
\ee
and the complete Hamiltonian is written as:
\be
{\cal H}={\cal H}_{AT}^{(-)}+{\cal H}_{AT}^{(+)}+{\cal V}_{-1,1}\,.
\label{H_AT1}
\ee
In the case of the chain  junction, we introduce  an extra line of
spins at  $z=0$, which are connected
horizontally to the two subsystems through  the respective bulk couplings and
there is a two-spin interaction associated with  the junction in the vertical direction.
In the Hamiltonian limit, the  different terms in ${\cal H}_{AT}^{(\pm)}$
are extended up to $i=0$ and the junction involves a transverse-field term 
\be
\widetilde{\cal V}_0=-\widetilde{h}(\sigma_0^x + \tau_0^x +
\epsilon_{\rm int} \sigma_0^x \tau_0^x)\,.
\label{V_h}
\ee
The ladder  and  chain  defects are transformed into each other through duality.
In the following, we study the ladder problem as defined in Eq.~(\ref{H_AT1}). 

\subsection{Basic quantities}

\subsubsection{Matrix elements}

We are interested in the local critical behavior of the system;
in particular, we want to determine the anomalous dimensions associated with the interface, $x_m^{\rm int}$ for the magnetization density ${\cal O}_m(0)$ and  $x_e^{\rm int}$ for the  energy density  ${\cal O}_e(0)$. These can be deduced from the  finite-size scaling of the singular
part of the corresponding matrix elements: 
\beqn
m_{\rm int}(L)&=&\langle 0 | \sigma^z(\pm 1) | 0 \rangle \sim L^{-x_m^{\rm int}}\,,\cr
e ^x_{\rm int}(L)&=&\langle 0 | \sigma^x(\pm 1) | 0 \rangle -e^x_{\rm int}\sim L^{-x_e^{\rm int}}\,,\cr
e ^z_{\rm int}(L)&=&\langle 0 | \sigma^z(-1)\sigma^z(1) | 0 \rangle -e^z_{\rm int}\sim L^{-x_e^{\rm int}}\,.
\label{matrix_e}
\eeqn
For the magnetization density, symmetry-breaking boundary conditions are needed. $| 0 \rangle$ is the ground state of the Hamiltonian in Eq.~(\ref{H_AT1}) and
$e^{x,z}_{\rm int}$ is the limiting value of the interface energy density in the
infinite system. We note that the two first matrix elements can be calculated on each side of the
interface and  there are two possible definitions for the energy density, $ e^x_{\rm int}$ and $e^z_{\rm int}$, corresponding to vertical and horizontal bonds in the classical model.

\subsubsection{Gaps}

These exponents can also be obtained by using  conformal
invariance.\cite{cardy87} The classical system composed of two semi-infinite planes coupled by one junction is mapped through the logarithmic transformation into two infinite  strips, each with a width $L/2$, coupled together by two parallel junctions at their boundaries and thus building a cylinder. In  the extreme anisotropic limit, a Hamiltonian ${\cal H}_{\rm cyl}$, similar to $\cal{H}$ in Eq.~(\ref{H_AT1}), is associated to the transfer matrix along the cylinder but with two junctions and periodic boundary conditions. For  a ladder defect, the two junctions are of the form given in Eq.~(\ref{V}), the first  between sites $i=-1$ and $i=1$ 
and the second between sites $i=-L/2$ and $i=L/2$. For a chain junction, the two junctions are of the form given in Eq.~(\ref{V_h}) and placed at $i=0$ and $i=L/2$.

In the cylinder  geometry, the first gap of ${\cal H}_{\rm cyl}$ scales as $1/L$  for a critical system, and the prefactor is proportional to the anomalous  dimension of the magnetization at the junction\cite{Cardy84}
\be
E_1-E_0=\frac{2\pi v_s}{L} x_m^{\rm int}\,.
\label{gap_exp}
\ee
Other local exponents are similarly related to higher gaps.

Before calculating the anomalous  dimensions numerically, we first consider the possible phase diagrams by studying the stability of the different fixed points. 

\subsection{Two identical subsystems}
\label{sec:id}
We start with the symmetrical model where $\epsilon^{(-)} = \epsilon^{(+)}=\epsilon$.
In this case there are three fixed points, located at $J=0$, $J=1$ and $J=\infty$,
and  corresponding  respectively to two disjoint semi-infinite systems (ordinary interface transition), to the homogeneous system (bulk transition), and to a system with an
ordered interface (extraordinary interface  transition).\cite{burkhardt81_rev,burkhardt81,diehl_diet_eisenr}

\subsubsection{Ordinary interface fixed point}
\label{sec:SI}
At the ordinary interface  fixed point the perturbation takes the form $J {\cal O}_m(-1){\cal O}_m(1)+J\epsilon_{\rm int} {\cal O}_p(-1){\cal O}_p(1)$. The first operator, involving the product of two surface magnetization operators, has the dimension 
\be
x_{\rm int}=x_m^{(-)}+x_m^{(+)}=2x_m^s\,,
\ee
and, thus, the scaling exponent of $J$ is
\be
y_{\rm int}=d_{\rm int} -x_{\rm int} =1-2x_m^s\,,
\ee
where $d_{\rm int}=d-1$ is the dimension of the interface.
This type of perturbation is irrelevant for $y_{\rm int}<0$, i.e., for $x_m^s>1/2$, which happens for $\epsilon>0$, whereas it is relevant for $\epsilon<0$. The marginality condition is satisfied for $\epsilon=0$, which is the Ising limit. The second operator, containing the product of two
surface polarization operators, has the dimension $\widetilde{x}_{\rm int}=2x_p^s=2$; therefore, this perturbation is always irrelevant.

\subsubsection{Bulk  fixed point}
The perturbation to the bulk  fixed point  introduced by the junction now takes  the form $\Delta {\cal O}_e(-1)+\widetilde{\Delta} {\cal O}_4(-1)$ where $\Delta=J-1$ and $\widetilde{\Delta}=J\epsilon_{\rm int}-\epsilon$. The dimension of the first operator is $x_{\rm int}=x_e$; thus, the scaling dimension of $\Delta$ is 
\be
y_{\rm int}=d_{\rm int} -x_e =d-1-x_e=\nu^{-1}-1\,.
\ee
This perturbation is relevant (irrelevant) for $\nu<1$ ($\nu>1$), i.e., for $\epsilon>0$ ($\epsilon<0$). The marginal situation corresponds once more to the Ising limit $\epsilon=0$.
The second operator $ {\cal O}_4$ has the scaling dimension $x_4=2$. It follows that $\widetilde{\Delta}$ has the scaling dimension $\widetilde{ y}_{\rm int}=-1$. Thus, the four-spin interface  perturbation is always irrelevant as for the ordinary interface fixed point.  

\subsubsection{Extraordinary interface fixed point}
\label{sec:OI}
The stability of this fixed point is related to that of the ordinary interface fixed point. Let us  consider  the chain junction in Eq.~(\ref{V_h}). The ordered
interface can be realized by setting the transverse field at the fixed point value $\widetilde{h}=0$. Under the duality transformation in Eq.~(\ref{dual}), the (weak) chain junction is transformed into a
(weak) ladder junction; consequently, to decide about the stability of the corresponding fixed
point, one can repeat  the argument of Sec.~\ref{sec:SI}.


\begin{figure}
\begin{center}
\vskip -1cm
\includegraphics[width=6cm]{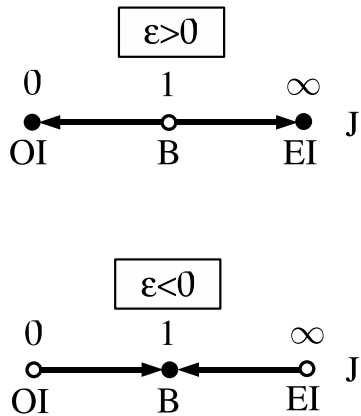}
\end{center}
\vskip -2cm
\caption{Schematic RG phase diagram at a ladder defect with coupling $J$ in the critical AT
model. The RG flow is different for different signs of the bulk four-spin coupling $\epsilon$. When $\epsilon>0$,  the bulk  fixed point (B) is unstable;  the flow is toward the ordinary interface  fixed point (OI) when $J<1$ and the extraordinary interface fixed point (EI) when $J>1$. When $\epsilon<0$, the flow is reversed and the bulk fixed point is always stable for $0<J<\infty$.}
\label{fig:phase1}
\end{figure}



\begin{figure}
\begin{center}
\vskip -1cm
\includegraphics[width=6cm]{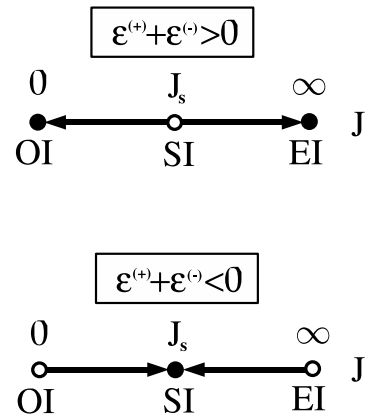}
\end{center}
\vskip -2cm
\caption{Schematic RG phase diagram at the interface  between two different critical  semi-infinite AT models with four-spin couplings $\epsilon^{(-)} < \epsilon^{(+)}$.
The interface coupling $J$ is of the ladder type. The direction of the RG flow depends on
the sign of $\epsilon^{(-)} + \epsilon^{(+)}$. When $\epsilon^{(-)} + \epsilon^{(+)}>0$, the special interface fixed point (SI) is unstable; the flow  is toward the ordinary interface  fixed point (OI) when $J<J_s$ and the extraordinary interface fixed point (EI) when $J>J_s$. When $\epsilon^{(-)} + \epsilon^{(+)}<0$, the flow is reversed and the special interface fixed point is always stable for $0<J<\infty$.}
\label{fig:phase2}
\end{figure}


\subsubsection{Renormalization group phase diagram}

Based on the stability analysis of the fixed points, the expected interface RG phase diagram
is depicted in Fig.~\ref{fig:phase1}. 

When $\epsilon<0$ and for any interface coupling $0<J<\infty$, the behavior at the interface is expected to be governed by the bulk fixed point. Then, the first gap in the spectrum of the conformal Hamiltonian ${\cal H}_{\rm cyl}$ has a $1/L$ dependence with a prefactor which, according to the gap-exponent relation in Eq.~(\ref{gap_exp}), is proportional to $x_m$. 

On the contrary, for $\epsilon>0$ the bulk fixed point is unstable. For weak couplings, $J<1$, the interface renormalizes to a cut and the critical behavior is the same as at a free surface. The first gap in the spectrum of ${\cal H}_{\rm cyl}$ for small $J$ can be estimated perturbatively, as in Sec.~\ref{sec:SI}. It behaves  as the product of the two surface magnetizations, vanishing as $\sim L^{-2x_m^s}$, which is faster than $1/L$  since $x_m^s>1/2$ according to Eq.~(\ref{surf-exp}). This  indicates that the system is asymptotically breaking into two pieces.  For strong couplings $J>1$, the interface remains ordered at the critical temperature and, through duality, the gap has also the size dependence $\sim  L^{-2x_m^s}$, which is faster than $1/L$.  It corresponds to a vanishing amplitude in Eq.~(\ref{gap_exp}) and, thus, to a vanishing interface magnetic exponent, a value which is linked with the local order at the critical point.
 
In the limit $\epsilon=0$, i.e., when the AT model becomes a system of two noninteracting  Ising models, the interface coupling $J$ is a marginal perturbation and the
local magnetization exponent is $J$ dependent,\cite{bariev79}
\be
x_m^{\rm int}(J)=\frac{2}{\pi^2} \arctan^2(1/J),\quad x_e^{\rm int}=1,\quad \epsilon=0\,.
\label{marg_0}
\ee
Similarly, for a chain junction,\cite{bariev79} the local magnetization exponent is $\widetilde{h}$ dependent,
\be
x_m^{\rm int}(\widetilde{h})=\frac{2}{\pi^2} \arctan^2(\widetilde{h}),\quad x_e^{\rm int}=1,\quad \epsilon=0\,.
\label{marg_1}
\ee
The marginal operator  is the local energy density, which keeps its
anomalous dimension $x_e^{\rm int}=1$, independently of the value of $J$ or $\widetilde{h}$.
We note that nonuniversal interface critical behavior at a defect plane can be found
in the three-dimensional $n$-vector model in the limit $n \to \infty$, which has been explicitly
calculated,\cite{eisenr_burkh}

\subsection{Two different subsystems}
\label{sec:nid}
If the two subsystems have different four-spin couplings $\epsilon^{(-)} < \epsilon^{(+)}$, one
can no longer define a bulk system fixed point. However, the ordinary and extraordinary  interface  fixed points still exist. The stability analysis of the ordinary interface fixed point can be performed along the lines of Sec.\ref{sec:SI}, leading to an interface exponent $y_i=1-x_m^{(-)}-x_m^{(+)}$. The ladder perturbation is irrelevant, i.e., the ordinary interface fixed point is stable (unstable) for $\epsilon^{(-)} + \epsilon^{(+)}>0$ ($<0$). Through duality, as described in Sec.\ref{sec:OI},
the same type of stability is expected to hold for the extraordinary interface fixed point, too. Consequently, the directions of the RG flows are analogous to the  case of identical subsystems in Fig.~\ref{fig:phase1}; just the role of the bulk fixed point is taken over by a new special interface fixed point, located at $J_s=O(1)$, which controls a special  transition. The expected RG phase diagram is given in Fig.~\ref{fig:phase2}.

The stability or instability of the special interface fixed point requires that the
scaling  dimension of the local energy-density operator satisfies 
\beqn
x_e^{\rm int}>1 \quad {\rm for} \quad \epsilon^{(-)} + \epsilon^{(+)}<0\,,\cr
x_e^{\rm int}<1 \quad {\rm for} \quad \epsilon^{(-)} + \epsilon^{(+)}>0
\label{stab}
\eeqn
at this fixed point.
 
In the borderline case, $\epsilon^{(-)} + \epsilon^{(+)}=0$, the perturbation is marginal at the ordinary and extraordinary fixed points. It is interesting to determine whether the interface remains marginal for any value of $J$, as it happens in the symmetric case. In the truly marginal case
(i) the local magnetization exponent is a continuous function of the coupling: $x_m^{\rm int}= x_m^{\rm int}(J)$ [as in Eqs.~(\ref{marg_0}) and~(\ref{marg_1})] and
(ii) the scaling dimension of the local energy-density operator has to remain constant: $x_e^{\rm int}=1$.


\begin{figure}
\begin{center}
\includegraphics[width=\columnwidth]{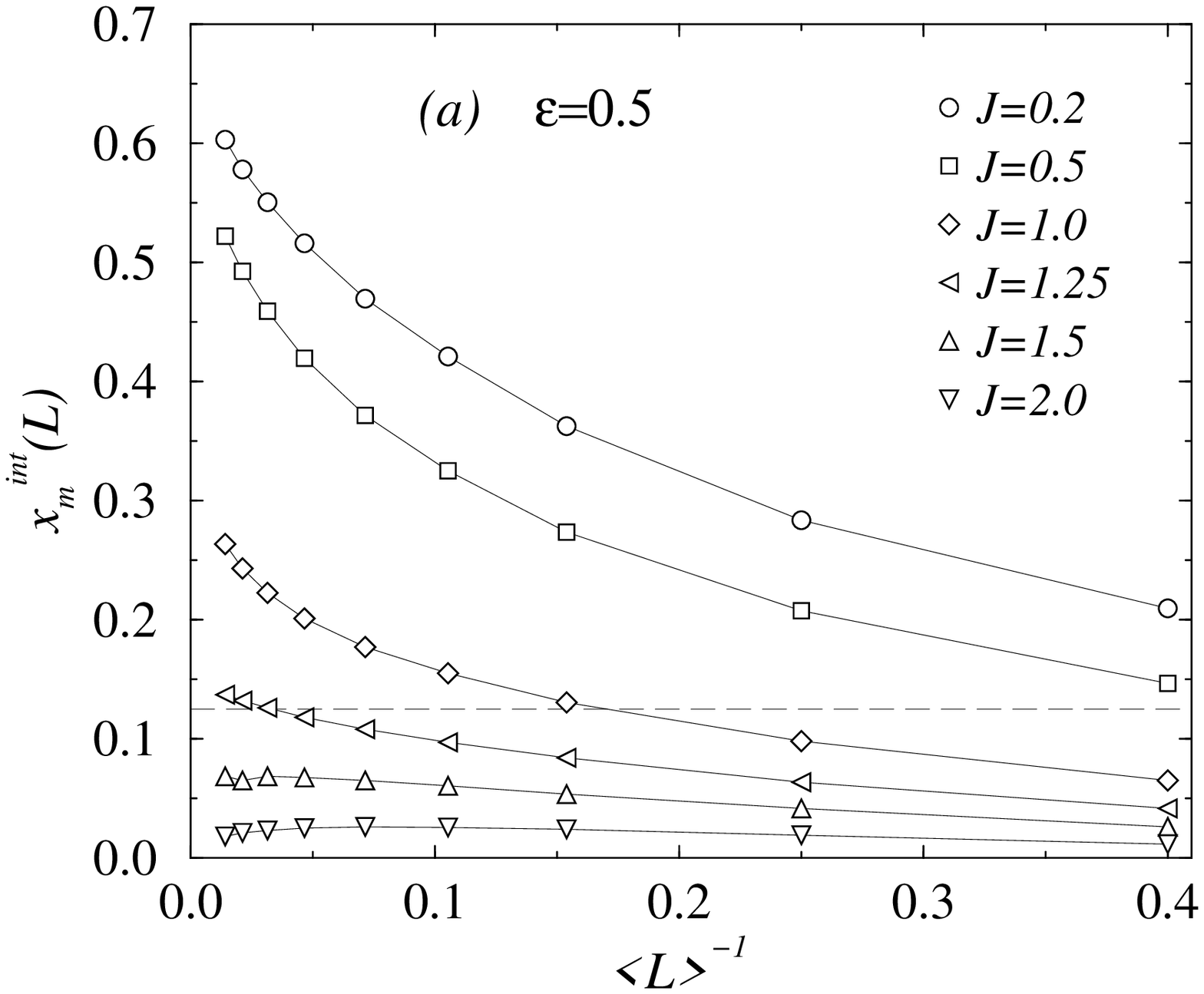}
\includegraphics[width=\columnwidth]{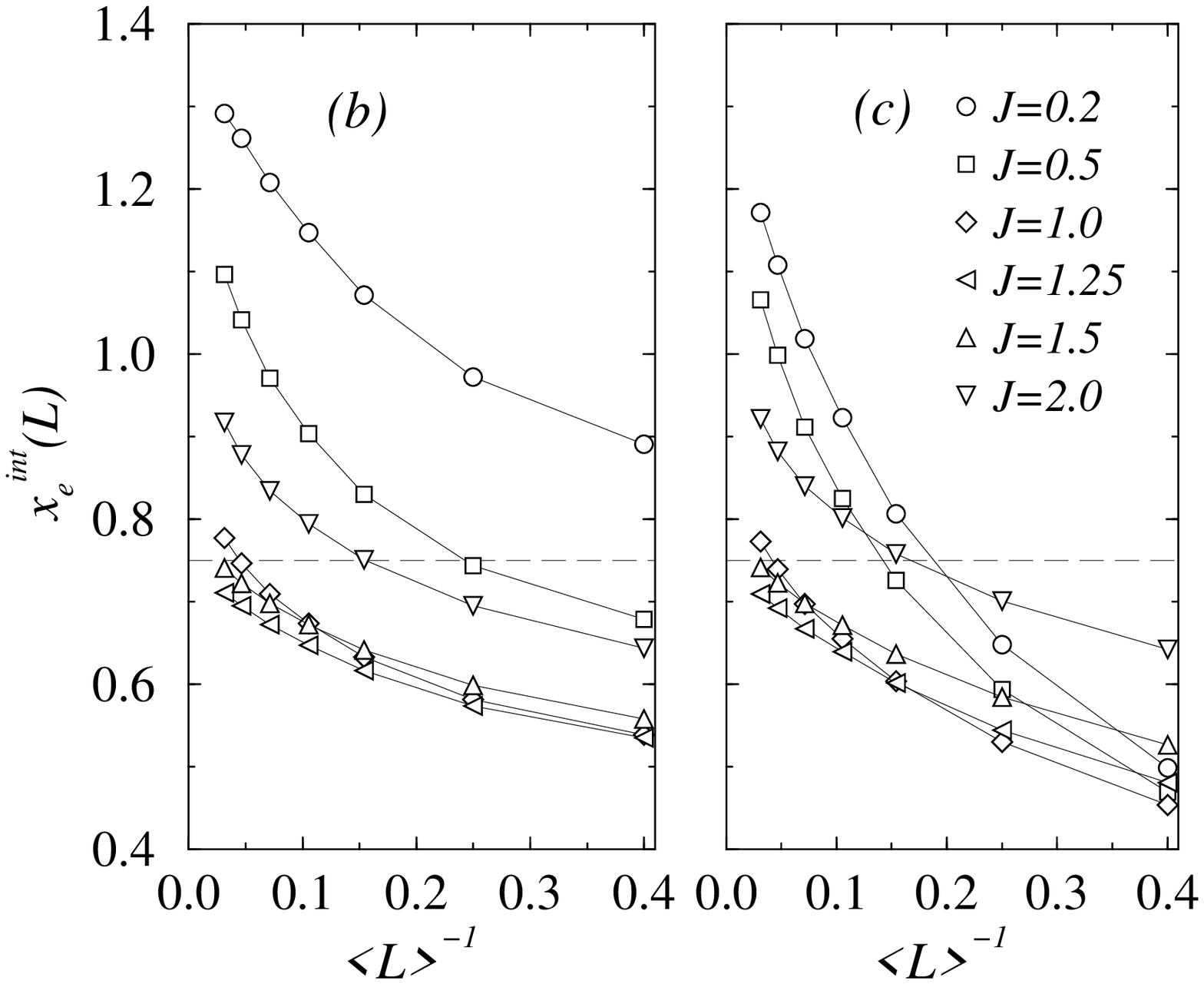}
\end{center}
\vskip 0cm
\caption{Interface critical behavior  between two identical Ashkin-Teller models with bulk
four-spin coupling  $\epsilon=0.5$ (relevant case).
The upper part gives the magnetization exponents and the lower part  the energy-density exponents deduced either 
from $e^x_{\rm int}$  on one side of the interface (left) or from $e^z_{\rm int}$ on the junction itself (right).}
\label{fig:iden05}
\end{figure}



\begin{figure}
\begin{center}
\includegraphics[width=\columnwidth]{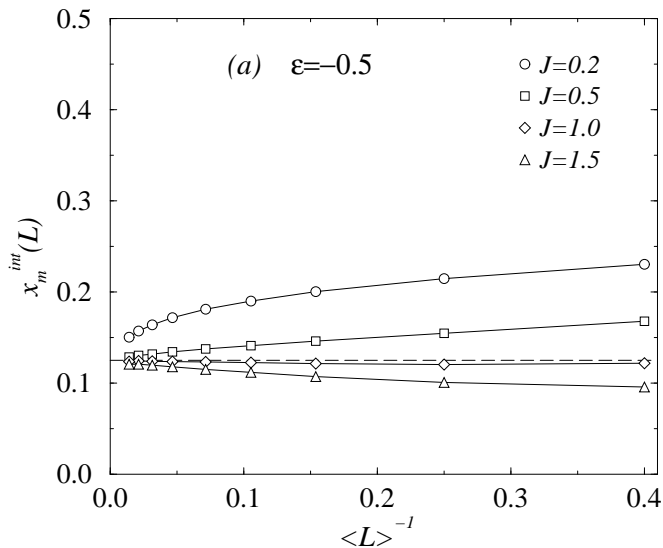}
\includegraphics[width=\columnwidth]{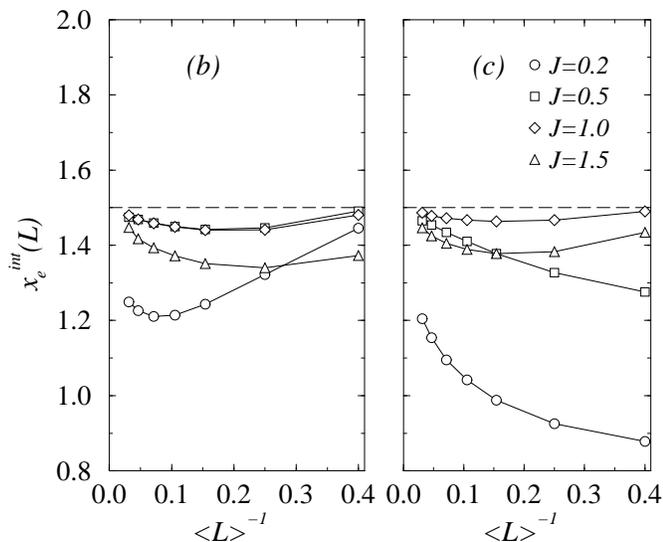}
\end{center}
\vskip 0cm
\caption{As in Fig.~\protect\ref{fig:iden05} for  the irrelevant case  $\epsilon=-0.5$.}
\label{fig:iden-05}
\end{figure}



\begin{figure}
\begin{center}
\includegraphics[width=\columnwidth]{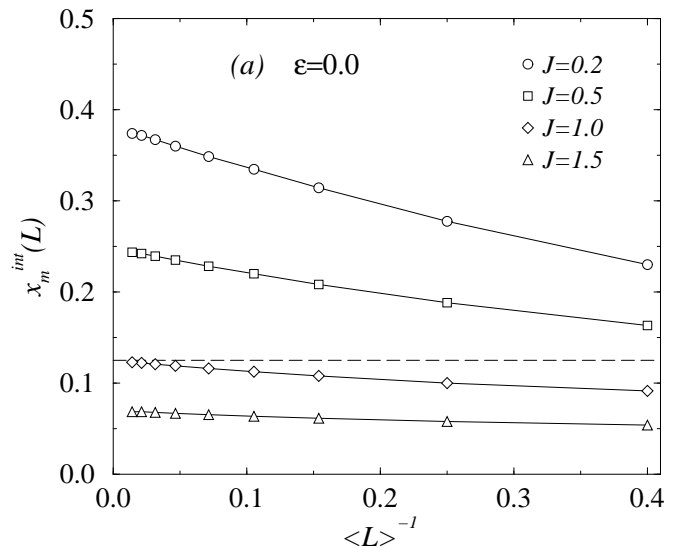}
\includegraphics[width=\columnwidth]{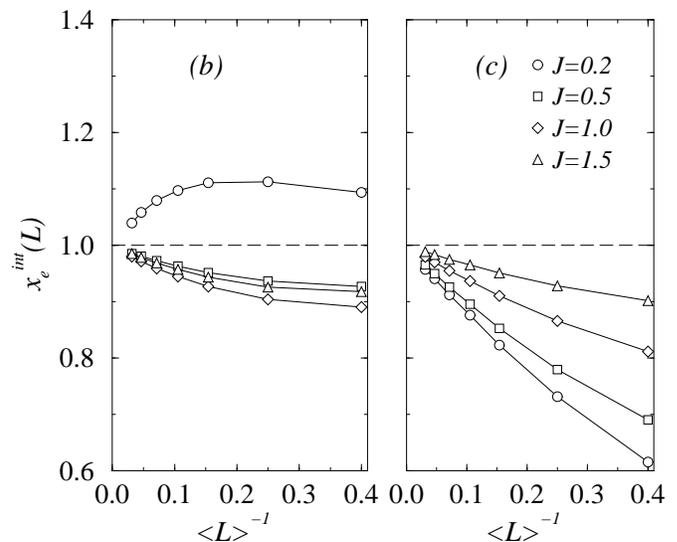}
\end{center}
\vskip 0cm
\caption{As in Fig.~\protect\ref{fig:iden05} for the marginal Ising limit $\epsilon=0$.}
\label{fig:iden0}
\end{figure}


\section{Numerical study}
\label{sec:num}
The calculation of the scaling dimensions associated with the interface, $x_m^{\rm int}$ and $x_e^{\rm int}$, is based on a finite-size scaling analysis of the matrix elements of the corresponding operators, as indicated in Eq.~(\ref{matrix_e}). The ground state
of the system, with a length $L$  for the two subsystems up to $86$ ($38$) for the magnetization (energy) exponent, has been  determined using the DMRG method.\cite{DMRG}  In order to obtain a good
accuracy, we have generally kept around $m=150$ states of the density matrix.

The magnetization density has been  calculated using symmetry-breaking boundary conditions, with the two types of spins  held fixed at both ends, $\sigma_{\pm L}^z=\tau_{\pm L}^z=+1$.  The magnetization density is determined on both sides of the interface when the system is asymmetric.

For the energy density, we eliminate the regular contribution to the ground-state expectation value in Eq.~(\ref{matrix_e}) by taking the difference of the values obtained for the systems with free and fixed boundary conditions. Since the sign of the singular part  generally changes when the boundary conditions are changed, a good precision can be obtained in this way.\cite{karevski96}
As indicated in Eq.~(\ref{matrix_e}), we calculate the energy density on the junction itself by taking the ground-state expectation value $\langle 0 | \sigma^z(-1)\sigma^z(1) | 0 \rangle$  and on both sides of the interface, with  $\langle 0 | \sigma^x(\pm 1) | 0\rangle$.

From the values  of the singular part of the  matrix element, say, $m_{\rm int}(L)\sim L^{- x_{m}^{\rm int}}$, at two different sizes, $L$ and $b L$, we deduce effective exponents through two-point fits:
\be
\frac{\ln m_{\rm int}(bL)-\ln m_{\rm int}(L)}{\ln b}=x_{m}^{\rm int}(L)\,.
\ee
In order to obtain the same numerical accuracy for the different points,  we  keep the ratio $b$ between neighboring sizes approximately constant. The effective exponents evolve  toward their  exact  values  when the mean size associated with the two-point fit, $\langle L\rangle=L(b+1)/2$, tends to infinity.

\subsection{Two identical subsystems}
\label{sec:num_id}

We first check  the validity of the phase diagrams given in Fig.~\ref{fig:phase1} for the interface between identical critical subsystems, the ladder defect in an otherwise homogeneous system.  We have studied three values of
the bulk four-spin coupling, $\epsilon=0.5,~-0.5$, and $0$, and calculated the
interface  magnetization and energy exponents for different values  of the interface coupling  $J$. The results are shown in 
Figs.~3--5.

When $\epsilon=0.5$ (Fig.~\ref{fig:iden05}) the perturbation is relevant and the bulk fixed point unstable. For  small values of the interface coupling, the flow is toward a free surface behavior. For $J=0.2$,  the effective exponents tend to their surface values, $x_m^{\rm int}=x_m^s=2/3$ and either $x_e^{\rm int}=x_e^s=2$ when the energy operator  is the surface energy operator of one subsystem ($ e^x_{\rm int}$)  or 
$x_e^{\rm int}=2x_m^s=4/3$ when the energy operator involves the surface magnetization operators of the two subsystems ($e^z_{\rm int}$).  The effective exponents converge slowly to $x_m^{\rm int}=0$ and $x_e^{\rm int}=2$, characteristic of an ordered interface, for the highest values of  $J$.  The interface exponents take the bulk values, $x_m^{\rm int}=1/8$ and $x_e^{\rm int}=3/2$,  for an intermediate value of $J$, between 1.25 and 1.5, where the flow is toward the (unstable in the $J$ direction) bulk fixed point.

For $\epsilon=-0.5$ the bulk fixed point is stable and the effective exponents in Fig.~\ref{fig:iden-05} approach the bulk values $x_m^{\rm int}=x_m=1/8$ and $x_e^{\rm int}=x_e=3/2$, independently of the value of the interface coupling. 

The Ising limit $\epsilon=0$ in Fig.~\ref{fig:iden0} is a truly marginal situation.  As expected, the  interface magnetization exponent  is continuously varying with $J$. The extrapolated values are in agreement with the exact results given in
Eq.~(\ref{marg_0}): $x_m^{\rm int}(J=0.2)=0.382$, $x_m^{\rm int}(J=0.5)=0.248$, and $x_m^{\rm int}(J=1.5)=0.07$, respectively. The interface energy exponent takes the bulk value $x_e^{\rm int}=1$,  which is necessary for a true marginal behavior at the line defect. 

\subsection{Two different subsystems}
\label{sec:num_nid}


\begin{figure}
\begin{center}
\includegraphics[width=\columnwidth]{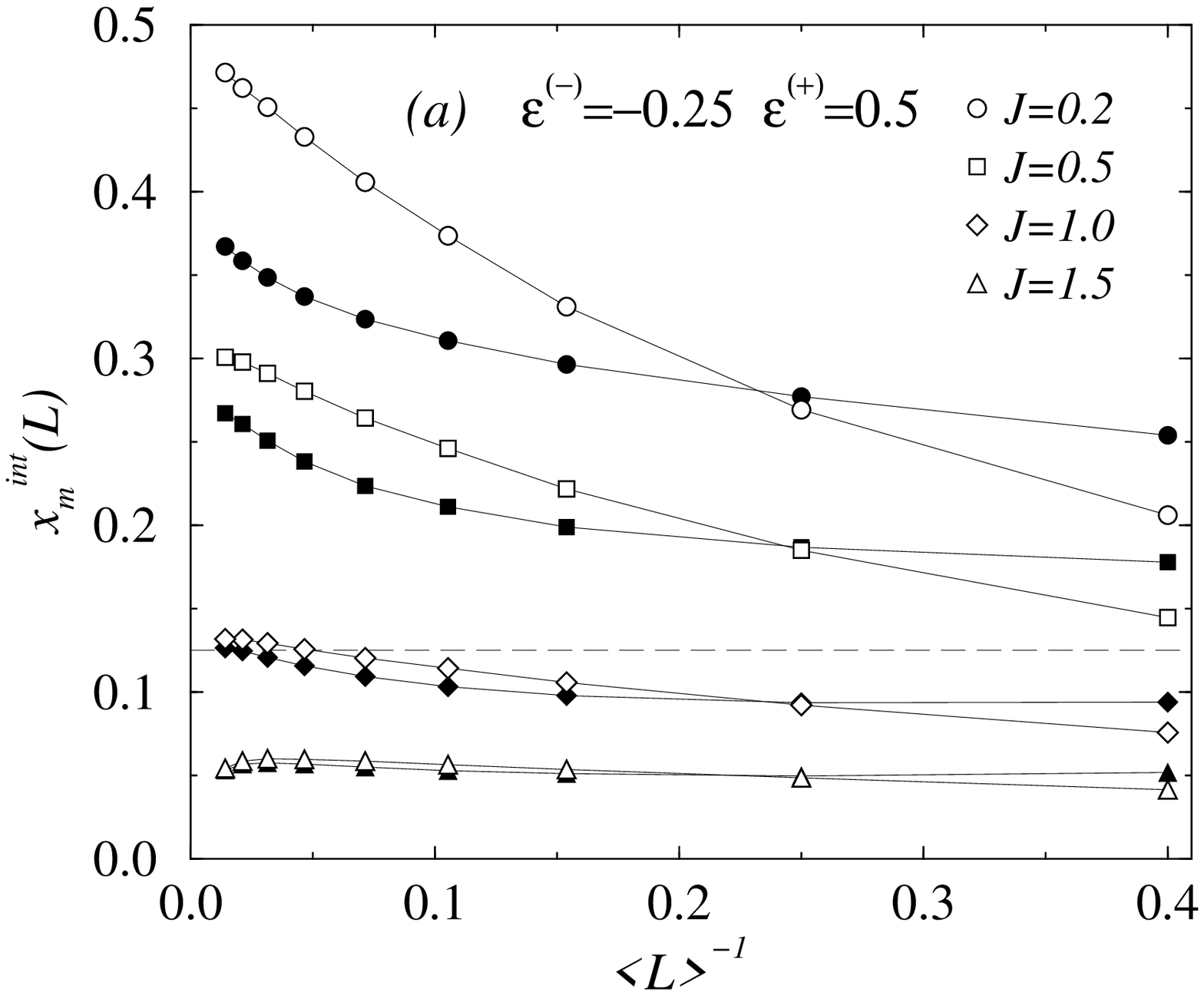}
\includegraphics[width=\columnwidth]{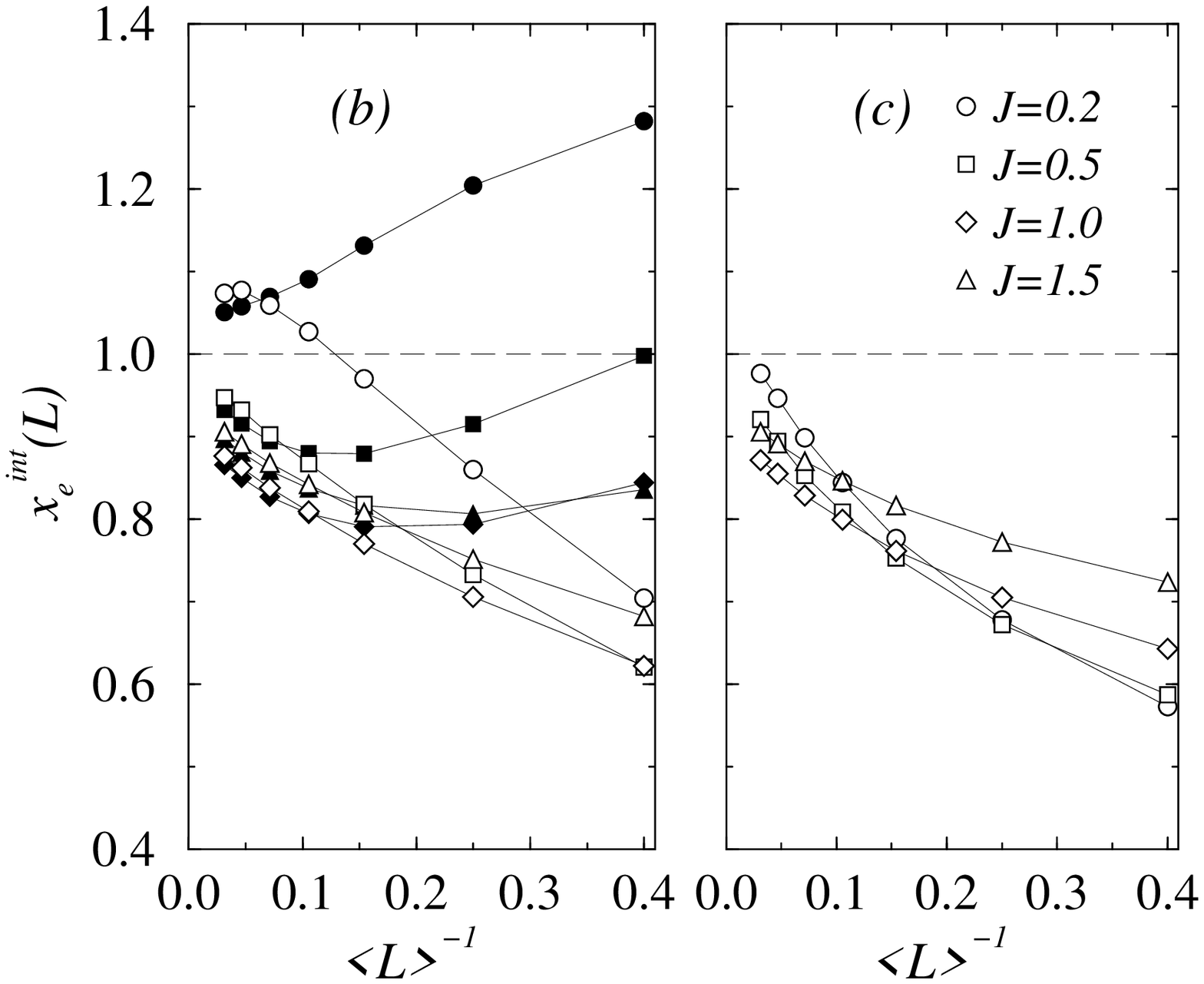}
\end{center}
\vskip 0cm
\caption{Interface critical behavior between two Ashkin-Teller models with
four-spin couplings $\epsilon^{(-)}=-0.25$ and $\epsilon^{(+)}=0.5$ (relevant case).
The upper part gives the magnetization exponents calculated either on the left (full symbols) or on the right (open symbols) of the interface.  The lower part gives the energy-density exponents deduced from $e^x_{\rm int}$  
 on the two sides of the interface (left) or from $e^z_{\rm int}$ on the junction itself (right).}
\label{fig:rel}
\end{figure}



\begin{figure}
\begin{center}
\includegraphics[width=\columnwidth]{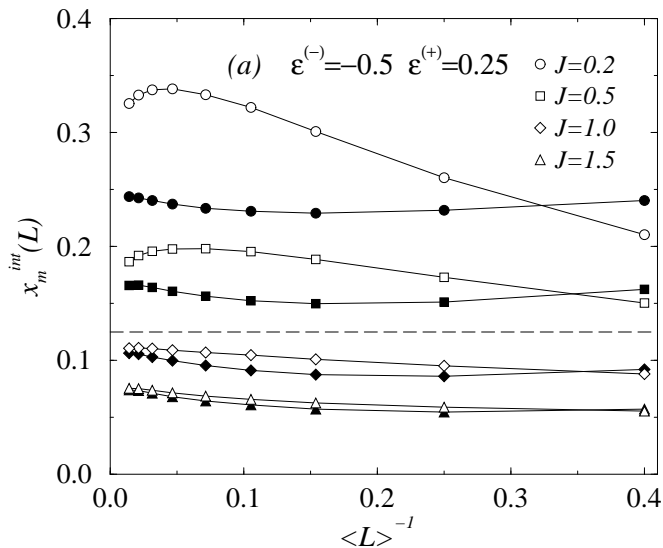}
\includegraphics[width=\columnwidth]{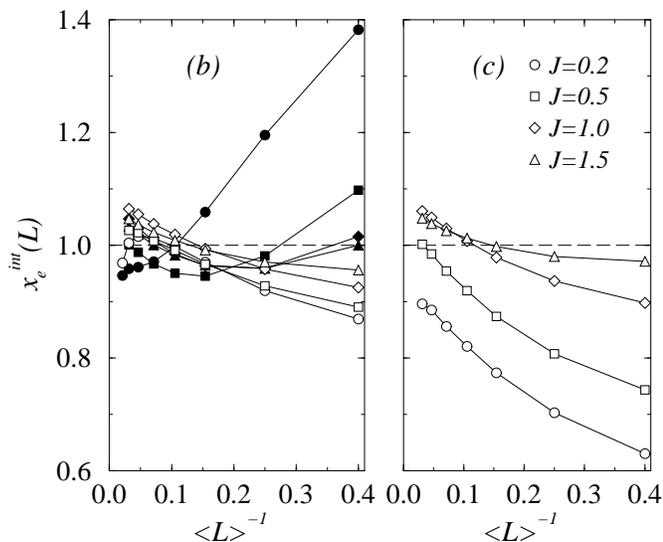}
\end{center}
\vskip 0cm
\caption{As in Fig.~\protect\ref{fig:rel} for $\epsilon^{(-)}=-0.5$ and $\epsilon^{(+)}=0.25$ (irrelevant case).}
\label{fig:irel}
\end{figure}



\begin{figure}
\begin{center}
\includegraphics[width=\columnwidth]{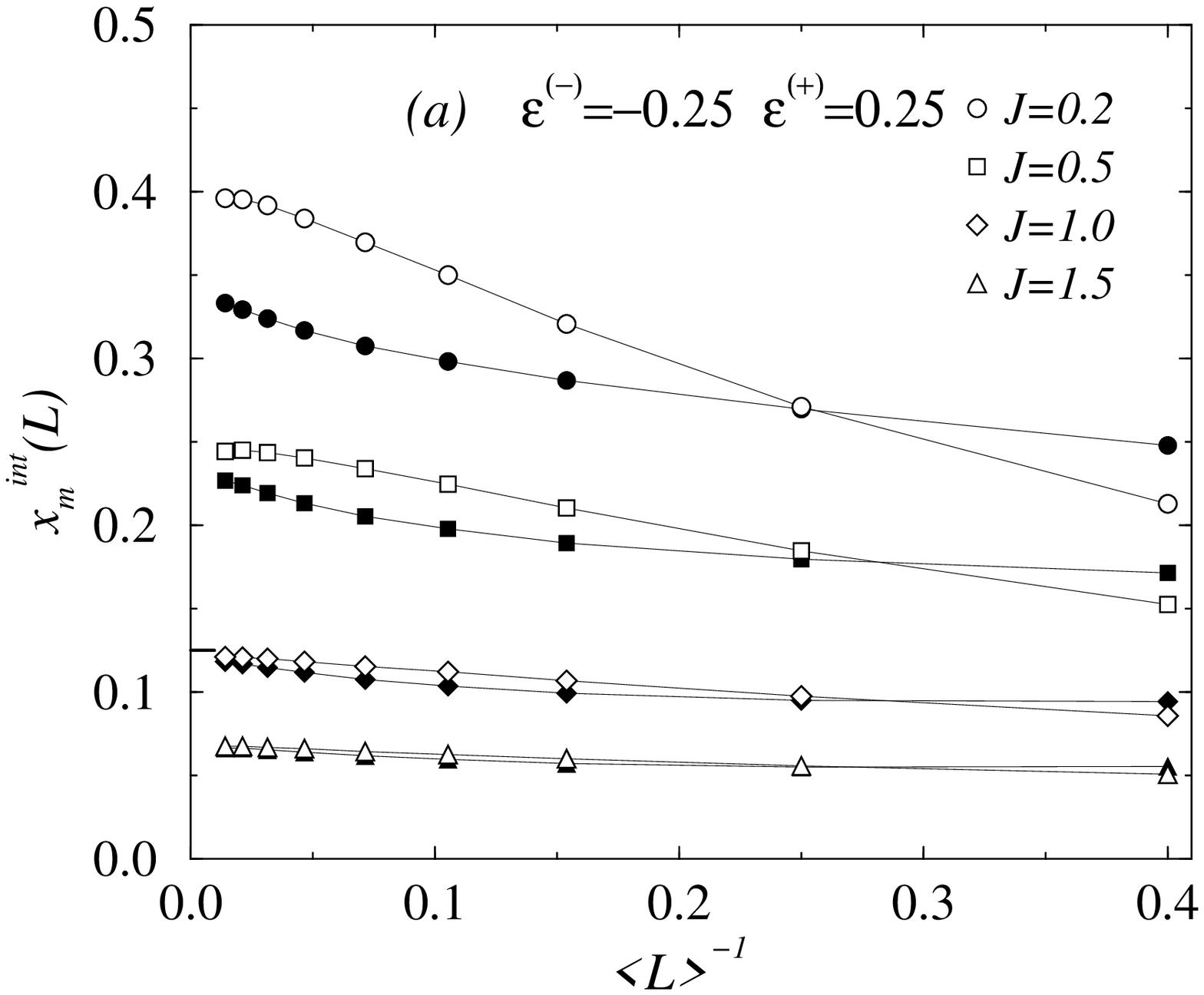}
\includegraphics[width=\columnwidth]{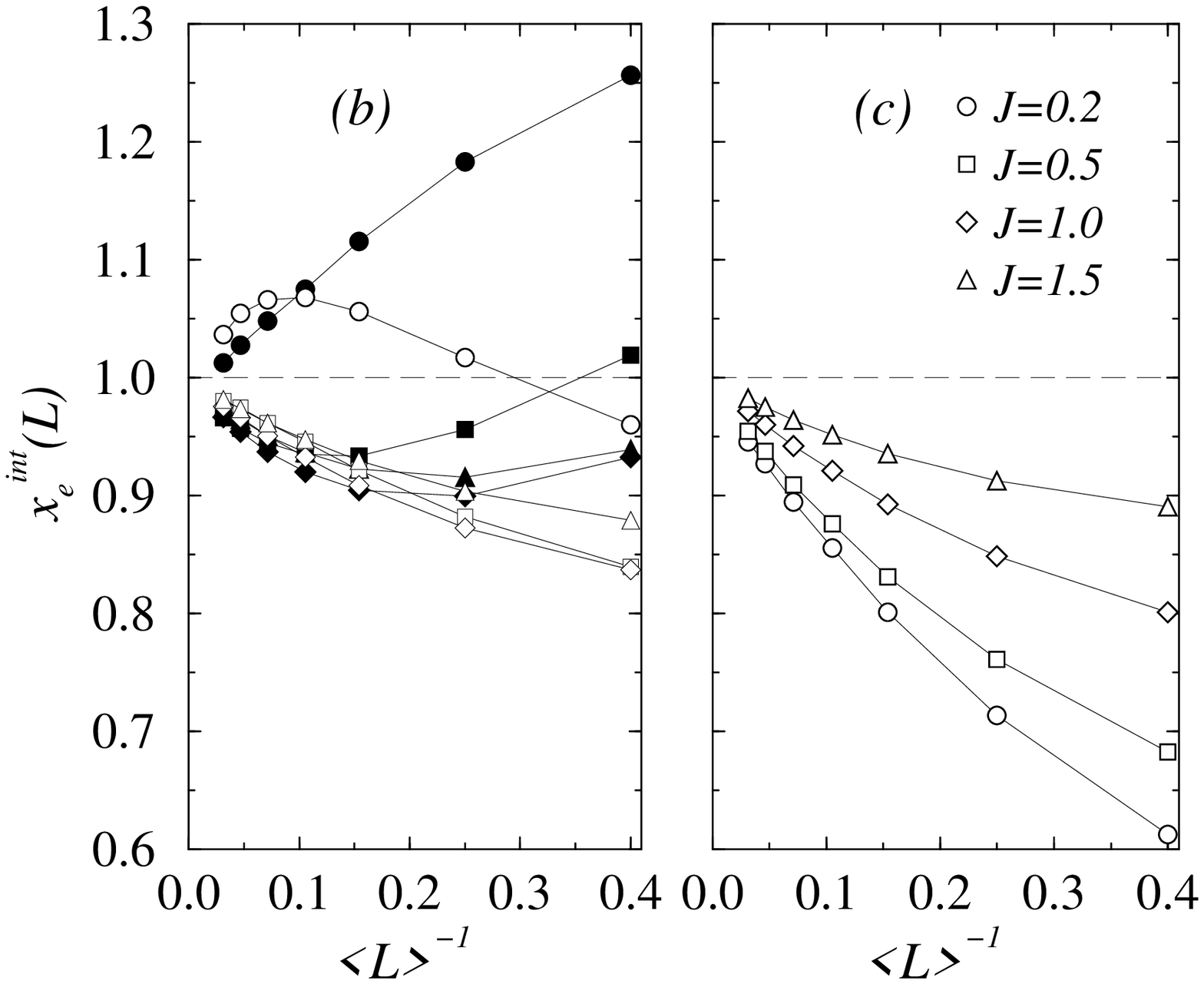}
\end{center}
\vskip 0cm
\caption{As in Fig.~\protect\ref{fig:rel} for the marginal case with  $\epsilon^{(-)}=-0.25$ and $\epsilon^{(+)}=0.25$.}
\label{fig:marg025}
\end{figure}



\begin{figure}
\begin{center}
\includegraphics[width=\columnwidth]{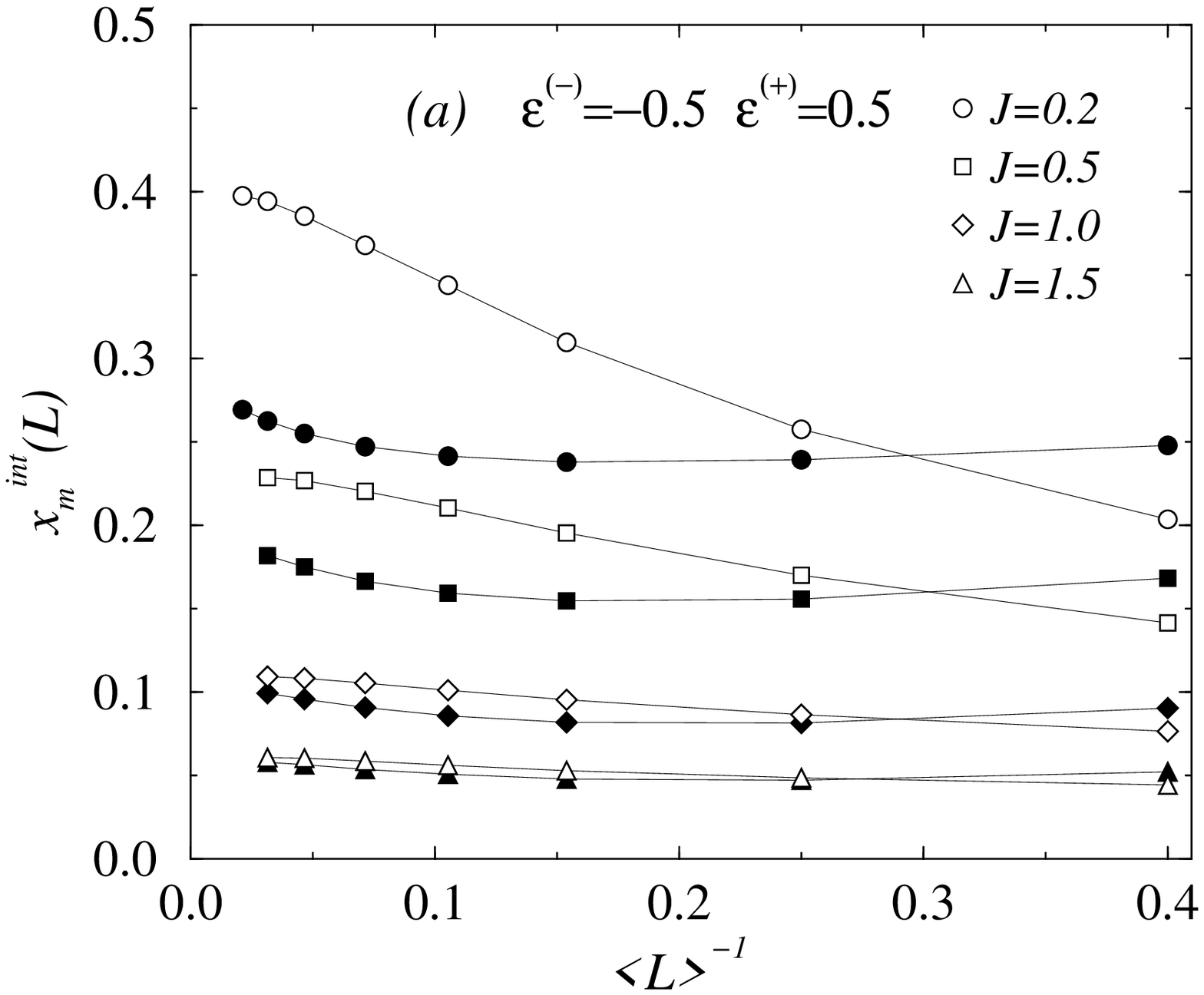}
\includegraphics[width=\columnwidth]{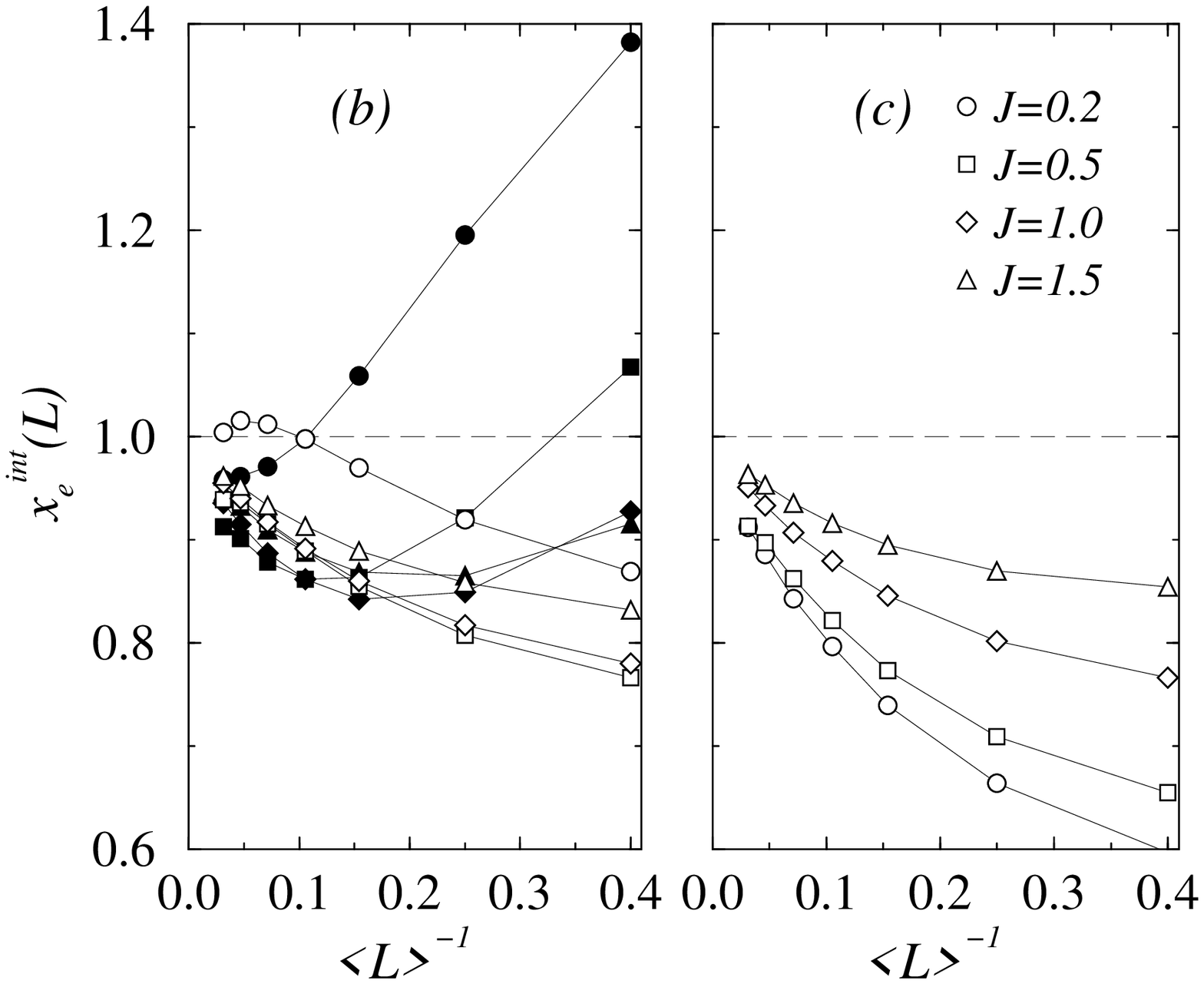}
\end{center}
\vskip 0cm
\caption{As in Fig.~\protect\ref{fig:rel} for  the marginal case with $\epsilon^{(-)}=-0.5$ and $\epsilon^{(+)}=0.5$.}
\label{fig:marg05}
\end{figure}


For an interface between two different subsystems, we start with the case where $\epsilon^{(-)}+\epsilon^{(+)}>0$, which corresponds to the
RG flow in the upper part of Fig.~\ref{fig:phase2}.  

The results obtained for the magnetization (energy) density exponents when $\epsilon^{(-)}=-0.25$
and $\epsilon^{(+)}=0.5$ are presented in the upper (lower) part of Fig.~\ref{fig:rel}. In accordance with the RG phase diagram, for small $J$ ($J=0.2$ and $0.5$), the effective interface exponents slowly approach the surface magnetization exponent of the right subsystem $x_m^s=2/3$, whereas in the other limit ($J=1.5$) they seem to converge to zero. According to the numerical results, the special interface transition takes place at $J_s \approx 1$ where the magnetization exponent  is close to $x_m=1/8$. 

The energy-density exponents shown in the lower part of Fig.~\ref{fig:rel} are greater for small and large values of $J$ than for $J=1.0\sim J_s$. This behavior is expected  since  $x_e^s=2$ at the ordinary and extraordinary transitions, whereas the instability of the special interface fixed point requires $x_e^{\rm int}<1$.

When  $\epsilon^{(-)}+\epsilon^{(+)}<0$, we are in the situation sketched in the  lower part of Fig.~\ref{fig:phase2} which was tested  for $\epsilon^{(-)}=-0.5$ and $\epsilon^{(+)}=0.25$. The numerical results  are presented in Fig.~\ref{fig:irel}.

Here, too, the crossover effects are quite strong for the effective magnetic exponents shown in the upper part of the figure. For a small coupling  $J=0.2$, the effective exponents remain close to the surface magnetization exponent of the $\epsilon^{(-)}=-0.5$ model, $x_m^s=1/3$, with a tendency to decrease at the largest sizes. The value of the coupling $J_s$ at the special interface fixed point is slightly higher than 1, where the  effective magnetization exponents have the smallest finite-size corrections and the extrapolated value is a little below $x_m=1/8$.

The stability of the fixed point  of the special interface  transition  is related to  the value
of the energy-density exponent  $x_e^{\rm int}$, which is shown in the lower part of Fig.~\ref{fig:irel}.
Except for $J=0.2$, the effective exponents extrapolates to values larger than 1,  in agreement
with the stability analysis in Eq.~(\ref{stab}). For $J=1.0$, one obtains $x_e^{\rm int}=1.10(2)$, which corresponds to a crossover exponent  $y_i=-0.10(2)$. This small (negative) value of the crossover exponent   explains  the slow convergence of the effective
magnetization exponents in the upper part of Fig.~\ref{fig:irel}.

For the marginal situation  where $\epsilon^{(-)}+\epsilon^{(+)}=0$, we considered
two cases,  $\epsilon^{(+)}=0.25$ and $0.5$. The results are shown in Figs.~\ref{fig:marg025} and~\ref{fig:marg05}, respectively. 

The magnetic exponents seem to vary continuously with $J$, without  evidence of crossover effects at  large sizes. The possibility that this system is  truly marginal  is supported by the behavior of the effective energy exponents  which, whatever the value of $J$,  extrapolate to a value compatible with $x_e^{\rm int}=1$. 

\section{Discussion}
\label{sec:disc}

One interesting feature of the interface critical behavior in the AT model is that the local critical
exponents are continuously varying with the strength of the junction when the sum of the four-spin couplings vanishes, even in the asymmetric case. Here, we discuss the possible origin of this truly marginal behavior.
 
We  consider a somewhat different  setting, where the  system is semi-infinite and consists of two subsystems with the shape of corners, $-\infty<z<0, 0<y<\infty$ and $0<z<\infty, 0<y<\infty$, 
connected by a chain junction along the sides at $z=0$. We are interested in the behavior of the generalized corner exponent  $x_m^c$,  measured at $y=z=0$.  Under a logarithmic conformal mapping,  the critical semi-infinite system  is transformed into a strip with open boundaries at $i=\pm L/2$ and a chain junction at $i=0$. In the Hamiltonian limit,  the strip Hamiltonian  ${\cal H}_{\rm str}$ involves a transverse field  $\widetilde{h}$  at $i=0$, whereas the two ends of the chain are
free. In the following, we calculate the first gap $\Delta E(\widetilde{h})$ of ${\cal H}_{\rm str}$, perturbatively for a small transverse field, and deduce the local scaling dimension $x_m^c$ 
through the gap-exponent relation of  Eq.~(\ref{gap_exp}), where $2\pi$ has to be replaced by $\pi$, the actual angle  in the mapping of the semi-infinite system. To calculate the gap,  we first  perform the duality transformation in Eq.~(\ref{dual}). The transformed chain has fixed boundary spins at $i=\pm L/2$ and  a (weak) defect coupling of strength $\widetilde{h}$ at $i=0$. The first gap is given by the difference of the ground-state energies with antiparallel  and parallel  boundary conditions: $ \Delta E(\widetilde{h})=E_0^{\uparrow \downarrow }-E_0^{\uparrow \uparrow }$. Actually,  antiparallel boundary conditions are applied  to one type of spin variables, say, $\sigma$, while parallel boundary conditions are always applied to the $\tau$ spin variables. To leading order, only the $\sigma$ spin variables contribute to the difference of the ground-state energies and we obtain
\beqn
\Delta E(\widetilde{h})&=&2m_s^{(-)}(L/2)m_s^{(+)}(L/2)\widetilde{h}\cr
&=&2 a^{(-)}a^{(+)}(L/2)^{-x_m^{(-)}-x_m^{(+)}}\widetilde{h}\;,
\label{pert}
\eeqn
where $m_s^{(\pm)}(L/2)=a^{(\pm)} (L/2)^{-x_m^{(\pm)}}$ is the surface magnetization in a  critical
chain with length $L/2$, when the spin of the same type on the other end is fixed in the up state. The second-order term vanishes since even contributions to $E_0^{\uparrow \uparrow }$ and to
$E_0^{\uparrow \downarrow }$ are exactly the same. Only  odd powers of $\widetilde{h}$
are present. 

The leading behavior of the gap in Eq.~(\ref{pert})  depends on the value of $x_i=x_m^{(-)}+x_m^{(+)}$. For $x_i>1$, the first gap vanishes faster than $1/L$; thus,  $x_m^c=0$ and the junction is ordered. This happens for $\epsilon^{(-)}+\epsilon^{(+)}>0$ and corresponds to the upper part of Fig.~\ref{fig:phase2}. On the contrary, for $x_i<1$, the gap has a decay slower than
$1/L$; thus, according to Eq.~(\ref{gap_exp}), the interface exponent $x_m^{\rm int}$  is formally divergent to
leading order of the perturbational calculation. This indicates that the extraordinary interface fixed point with $\widetilde{h}=0$ is unstable, a situation which corresponds to the lower part of Fig.~\ref{fig:phase2}. In the marginal case $x_i=1$, up to first
order in $\widetilde{h}$, the local exponent has the variation 
\be
x_m^c=4\widetilde{h}\frac{a^{(-)}a^{(+)}}{\pi v_s}+O(\widetilde{h}^3)\,,
\ee
where the coefficients $a^{(\pm)}$  are  $O(1)$. 

In the Ising limit  $\epsilon^{(\pm)}=0$,
$a^{(\pm)}=1$ and, with the parametrization chosen for  the quantum Hamiltonian, $v_s=2$. Thus we obtain $x_m^c(\widetilde{h})=\frac{2\widetilde{h}}{\pi}+O(\widetilde{h}^3)$, which is
the leading contribution to  the exact result:\cite{HPS89} $x_m^c(\widetilde{h})=1-\frac{2}{\pi}\arctan(\widetilde{h}^{-1})$.

In the asymmetric marginal case, $\epsilon^{(-)}+\epsilon^{(+)}=0$, we also have  a
continuous variation of the leading contribution to $x_m^c$ with $\widetilde{h}$. We also expect a
truly marginal local critical behavior in this case. This  assumption is supported by
the fact that the second-order term of the expansion is vanishing due to symmetry. In the  marginally relevant or irrelevant cases, the second-order term of the expansion is usually diverging as $\log L$,\cite{igloi91} however, for a marginally irrelevant perturbation, the singular terms are expected to sum up to a regular contribution. In our case,  we expect a truly marginal behavior and continuously varying local scaling exponents also  for the ladder junction studied numerically in Sec.~\ref{sec:num}.

There are other systems from  which similar  composite  critical systems can be built and for which a truly marginal interface critical behavior could be obtained.  Let us mention the 2D $XY$ model  with different temperatures, say, $T^{(-)}$ and $T^{(+)}$, both lower than  the Kosterlitz-Thouless temperature. Another example is the $XXZ$ chain with different anisotropies on the two sides of the junction. One may notice that the AT Hamiltonian can be transformed into  a staggered $XXZ$ model through a duality transformation of the $\tau$ spins followed by a duality transformation on all the spins.\cite{KdNK81} Finally, let us mention the Potts model in the Fortuin-Kasteleyn representation, for which  the number of states  $q$ becomes a continuous parameter. Taking two subsystems with $q$ states on one side and $4-q$ states on the other, since $x_m^s(q)=1-(2/\pi)\arccos(\sqrt{q}/2)$,\cite{cardy84b} one has $x_m^s(q)+x_m^s(4-q)=1$. It follows  that the junction is a marginal perturbation at small coupling. In this case, however, the central charges of the
conformal field theory at the two sides of the junction are different; therefore, it needs
further investigations to decide if the perturbation remains truly marginal for any coupling strength, as observed in the symmetric Ising limit $q=2$. 

\begin{acknowledgments}
  This work has been supported by the National Office of Research and
Technology under Grant No. ASEP1111, by the Hungarian National Research Fund 
under Grants No. OTKA TO48721, No. K62588, and No. MO45596. F.I. thanks Universit\'e Henri Poincar\'e for hospitality. The Laboratoire de Physique des Mat\'eriaux is Unit\'e Mixte de Recherche CNRS No. 7556.
\end{acknowledgments}

\end{document}